\documentclass[aps,reprint,amsmath,amssymb]{revtex4-2}
\usepackage{graphics}
\usepackage{svg}
\usepackage{subfigure}
\usepackage{epsfig}
\usepackage{epsf,epic}
\usepackage{dcolumn}
\usepackage{color}
\usepackage{hyperref}
\usepackage{marvosym}
\usepackage{subfigure}
\usepackage{amsmath}
\usepackage{mathrsfs}
\usepackage{amsfonts}
\usepackage{amssymb}
\usepackage{wrapfig}
\usepackage{textcomp}   
\usepackage{fixltx2e}   
\usepackage[T1]{fontenc}   
\usepackage{pstricks}
\usepackage{multirow}
\usepackage{bm}
\usepackage{threeparttable}%
\graphicspath{{./figs/}}
\usepackage{dcolumn}
\newcommand{\etal}{\textit{et al.\ }}
\newcommand{\ie}{\textit{i.e.\ }}

\begin{document}
	\title{First-principles study  of infrared, Raman, piezoelectric and elastic  properties of Mg-IV-N\textsubscript{2} (IV = Ge, Si, Sn)}
	\author{Sarker Md. Sadman}
	\author{Walter R. L. Lambrecht}\email{walter.lambrecht@case.edu}
	\affiliation{Department of Physics, Case Western Reserve University, 10900 Euclid Avenue, Cleveland, Ohio 44106-7079, USA}
	\begin{abstract} 
          Mg-IV-N\textsubscript{2} compounds with IV=Si, Ge, Sn are ultra-wide band gap semiconductors with various potential electronic and optoelectronic applications.
           They share the  \textit{Pna}2\textsubscript{1} space group  crystal structure. Here we present Density Function Perturbation Theory (DFPT) calculations of  the vibrational modes  of these materials. We focus  on the vibrational modes at the zone center to establish the relation between vibrational modes and their corresponding point-group symmetries, which determine the Raman and infrared spectra but also report the full Brillouin zone phonon dispersions and density of states. We also  determine the piezoelectric tensor and the elastic compliance  tensor. 
	\end{abstract}
		\maketitle
	\section{Introduction}
        The group-III nitride semiconductors GaN and InN have gained an important place in electronic and optoelectronic device technologies. Their large direct band gaps  have enabled blue and white light emitting diodes (LED) and lasers but also advances in high-power and high-frequency electronic devices. In moving these technologies to even higher band gaps, so-called ultra-wide-band-gap (UWBG) semiconductors, for deep UV optoelectornics and high-power applications, the problems with doping AlN and high Al content Al$_x$Ga$_{1-x}$N alloys have become a barrier toward progress. While there have been  recent breakthroughs on this problem\cite{ahmad2021,Ahmad2022}, an alternative approach is to broaden the chemical space by considering the family of heterovalent II-IV-N$_2$ semiconductors.
        
In these materials each N is ideally surrounded by two group II and two group IV elements instead of four group III elements. Based on the wurtzite parent structure, an ordered arrangement of the group-II and group-IV in respective sublattices leads to an orthorhombic structure with spacegroup $Pna2_1$ observed for almost all members of this family. This opens a vast array of new possibilities as the group II elements can be taken from the IIA (Be, Bg, Ca) or IIB (Zn, Cd, Hg) columns or even the VIIB column (Mn) and each of these can be combined with IV=Si, Ge, Sn. Furthermore, as these compounds typically allow for a certain degree of  cation antisite disorder\cite{Skachkov16,Skachkov16x,Lany2017,Cordell2021,Cordell2022} as well as different polytypic stacking maintaining the local octet rule\cite{Quayle15} are possible ($Pna2_1$ and $Pmc2_1$), the band gap can be further tuned by modifying the crystal structure. It has thus been realized that these compounds  complement the group-III nitrides and can be further starting points for alloying among them or with group-III nitrides or by creating designed heterostructures.
Recent reviews of the status of these materials can be found in \cite{Lambrechtbook,Lyu2019,Martinez17}.

Within this family, recent work has drawn attention to the Mg based compounds for UWBG semiconductor applications. Their growth in thin film form was recently established by Metalorganic Chemical Vapor Deposition (MOCVD)\cite{Hu2025-MgSi,Hu2025-MgGe} and synthesis  in polycrystalline powder form was reported earlier\cite{Bruls1999}. The anisotropic thermal expansion and other thermal properties of MgSiN$_2$ were  studied by Bruls \etal \cite{Bruls2000,Bruls2001}.  Previous computational work has already determined the phase stability and thermodynamic properties \cite{Arab2016}. The electronic band structure of both the Si and Ge based compounds was first studied by Basalaev \etal \cite{Basalaev2010,Basalaev2011}, and more recently using reliably predictive methods, such as the quasiparticle self-consistent $GW$ method \cite{Atchara16,Lyu2019mg} and hybrid functional \cite{Quirk14} calculations. Phonons have also been studied \cite{Pramchu17,Rasander17,Kaewmeechai2017}, but a full account of the Raman spectra and infrared spectra of the full family including MgSnN$_2$ with a consistent approach is not yet available. In this paper we present a comprehensive study of the vibrational modes, their symmetry labeling, the related infrared and  Raman spectra including their polarization dependence, LO-TO-splitting, Born effective charges and phonon dispersions and densities of states.

	
	\section{Computational Method} \label{sec:method}
	Our calculations are based on density functional theory (DFT)  and density functional perturbation theory (DFPT) \cite{Gonze1997,Gonze1997a,Hamann2005} as implemented in the {\sc Abinit} code \cite{Gonze2020}. These calculations were performed on relaxed-structures optimized with the generalized gradient appproximation (GGA) in the Perdew-Burke-Ernzerhof (PBE) exchange functional with the Projector Augmented Wave (PAW) method \cite{PAW,Torrent2008}. The metric approach proposed by Hamann \etal \cite{Hamann2005} is used to deal with strain derivatives.
        Mg [2s\textsuperscript{2}2p\textsuperscript{6}3s\textsuperscript{2}], Ge [3d\textsuperscript{10}4s\textsuperscript{2}4p\textsuperscript{2}], Si [3s\textsuperscript{2}3p\textsuperscript{2}], Sn [4d\textsuperscript{10}5s\textsuperscript{2}5p\textsuperscript{2}], N [2s\textsuperscript{2}2p\textsuperscript{3}] electrons were treated as valence electrons. The wave function cut-off was chosen to be 50 Ha. Relaxation of the atoms and the structural parameters were carried out with the tolerance of forces of 1.0$\times$10\textsuperscript{-6}Ha/Bohr.
        	The force constants  and derivatives of the total energy versus strain and static electric field required for the calculation of phonons, LO-TO splitting,  infra-red  spectra  and Raman tensors  are obtained from first-order perturbed  wave functions using Density Functional Perturbation Theory (DFPT) \cite{Gonze1997,Gonze1997a}.  The metric approach proposed by Hamann \etal \cite{Hamann2005} is used to deal with strain derivatives. To calculate Raman intensities we need third order derivatives of the total energy  as discussed in \cite{Veithen2005} and in the PAW method, this requires second order perturbed wave functions \cite{Miwa2011}. The tolerance of energy was chosen as 1.0$\times$10\textsuperscript{-10}Ha/Bohr. In order to obtain accurate first and second order changes in the wave functions,  the ground state wave function was converged to a tolerance of 1.0$\times$10\textsuperscript{-09} Ha/Bohr. For the Brillouin zone sampling an unshifted  grid of size 8$\times$6$\times$8 was chosen.

	\section{Results and discussion}

	\subsection{Crystal structure}\label{sec:crystal}
\begin{figure}[htbp]
			\includegraphics[width=\columnwidth]{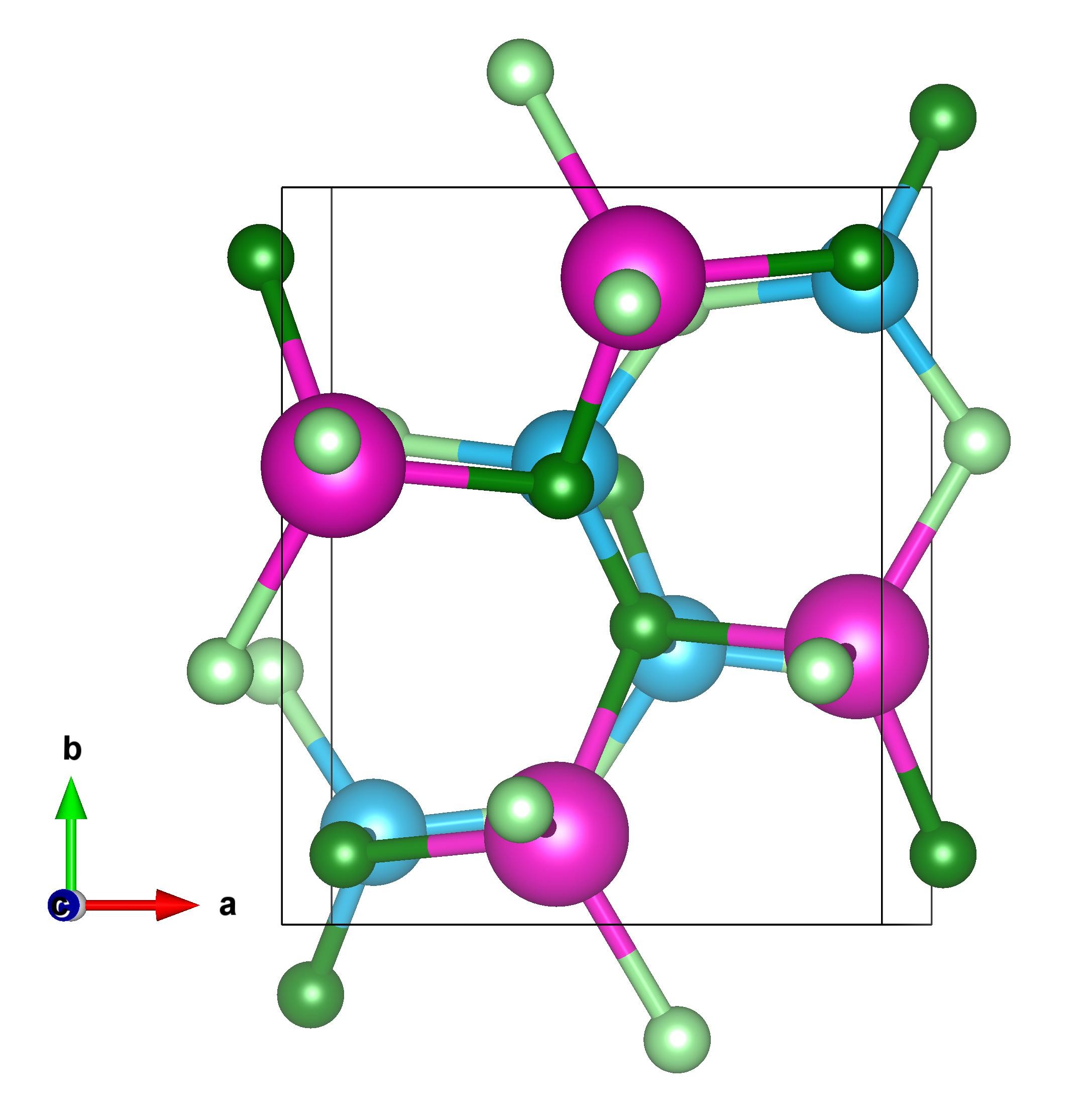}
			\caption{Crystal Structure of MgSiN\textsubscript{2} with Mg-atoms colored in pink, N-atoms in dark and light green and Si-atoms in light blue (left).} \label{crystal}
		\end{figure}

		The Mg-IV-N\textsubscript{2}(IV=Si,Ge,Sn) compounds share the same orthorhombic crystal structures with the $Pna2_{1}$
        spacegroup (number 33) and C\textsubscript{2v} point group symmetry.
        In this crystal structure each of the atoms occur in $4a$ Wyckoff positions. There are four formula units per cell or 16 atoms.  We started from
        the positions in Materials Project but subsequently relaxed it with the PBE functional.   Information related to crystal parameters, bond lengths, bond angles, and Wyckoff positions are provided in Table~\ref{latpara}. The structure is shown for MgSiN$_2$  in Fig. \ref{crystal} and is qualitatively similar in the other cases. We define an effective wurtzite lattice constant $a_w$ by $ab=2\sqrt{3}a_w$, \ie by setting the area of the orthorhombic basal plane cell equal to that of the equivalent wurtzite supercell. The  $c/a_w$ ratio is then seen to be almost the same 1.60 for all compounds. This $c/a$ is significantly lower than the ideal wurtzite $c/a=\sqrt{8/3}\approx 1.632$. The $a_w$ are close to their III-N parent compound AlN (3.11 \AA), GaN (3.19 \AA) and InN (3.54 \AA). The $b/a$ are all larger than the ideal $b/a=2/\sqrt{3}\approx 1.154$.
        Overall, the distortions from the ideal wurtzite structure are large  and decreasing from Si to Ge to Sn.  The Mg-N bond lengths are more or less constant while the IV-N bond lengths increase from Si to Ge to Sn. 
Note that the N$^{(1)}$ is located above the IV element and N$^{(2)}$ sits above the Mg. We thus see that the apical bond length (approximately along $c$ ) is shorter than the lateral bond length  (in the buckled hexagonal layer) for Mg while for the group-IV element  the two bond lengths are almost equal for Ge, slightly larger for the apical bond length for Si  and slightly smaller for Sn.

		\begin{table*}
		\caption{\large Lattice parameters, atomic postions, and bond lengths of Mg-IV-N\textsubscript{2} Crystal Structures \label{latpara}}
		\begin{ruledtabular}
			\begin{tabular}{cccc}
			Parameters & MgSiN\textsubscript{2} & MgGeN\textsubscript{2} & MgSnN\textsubscript{2} \\
			\hline   
			a & 5.31 & 5.56 & 5.97 \\
			b & 6.49 & 6.67 & 6.94 \\
			c & 5.03 & 5.23 & 5.54 \\
			$a_w$ & 3.154 & 3.272 & 3.458 \\
			$c/a_w$ & 1.595 & 1.599 & 1.602 \\
			$b/a$ & 1.22 & 1.20 & 1.16 \\
			\hline
			4a Mg & (0.08477, 0.62276, 0.98848) & (0.08514, 0.62342, 1.00034) & (0.08375, 0.62459, 0.99330) \\
			4a Si & (0.06967, 0.12541, 0.00012) & - & - \\
			4a Ge & - & (0.07361, 0.12578, 0.00739) & - \\
			4a Sn & - & - & (0.08312, 0.12579, 0.99585) \\
			4a N\textsuperscript{(1)}& (0.04788, 0.09500, 0.34646) & (0.06058, 0.10745, 0.36747) & (0.07819, 0.12246, 0.37482) \\
			4a N\textsuperscript{(2)}& (0.11001, 0.65601, 0.41096) & (0.10010, 0.64303, 0.40480) & (0.08512, 0.62713, 0.37702) \\
			\hline
			Mg-N\textsuperscript{(1)} & 2.14 & 2.12 & 2.12 \\
			Mg-N\textsuperscript{(2)} & 2.09 & 2.09 & 2.11 \\
			Si-N\textsuperscript{(1)} & 1.77 & - & - \\
			Si-N\textsuperscript{(2)} & 1.76 & - & - \\
			Ge-N\textsuperscript{(1)} & - & 1.89 & - \\
			Ge-N\textsuperscript{(2)} & - & 1.89 & - \\
			Sn-N\textsuperscript{(1)} & - & - & 2.09 \\
			Sn-N\textsuperscript{(2)} & - & - & 2.10 \\
						\end{tabular}
		\end{ruledtabular}
		\end{table*}

	        \subsection{Group theoretical analysis}\label{secgroup}
                Since the crystal structure has  16 atoms per unit cell, 48 vibrational modes are present. At $\Gamma$  the three modes with the lowest energies correspond to the collective translation motion of the crystal and  have zero energy. At general wave vector, they correspond to the acoustic modes.  The remaining 45 modes are optical modes which at $\Gamma$ can be labeled  in terms of the C\textsubscript{2v} point group irreducible representations.

                The character table shows that $A_1$, $B_1$, $B_2$ correspond to $z$, $x$, $y$ respectively, which are infrared active and exhibit LO-TO splitting and and $A_2$ corresponds to $xy$. All  modes are Raman active. The 4a Wyckoff position has no symmetry at all and thus has characters, 4,0,0,0 which decomposes into $A_1+A_2+B_1+B_2$. Because we have 4 types of atoms and each can move in three directions there are 12 modes of each irreducible representation but the  translation of the crystal in  $x,y,z$ corresponds to $B_1$, $B_2$, $A_1$.	Thus, the total optical modes of vibrations can be expressed as $\Gamma_o = 11A_1 + 12A_2 + 11B_1 + 11B_2$.
	
		\begin{table}
		\caption{Character Table of the group C\textsubscript{2v}(mm2).\label{character}}
		\begin{ruledtabular}
			\begin{tabular}{ccccccc}
				Modes & E & \( \sigma_v(xz) \) & \( C_2 \) & \( \sigma_v(yz) \) &  \multicolumn{2}{c}{functions}  \\
				\hline
				\( A_{1} \) & 1 & 1 & 1 & 1 & z & $x^2, y^2, z^2$ \\
				\( A_{2} \) & 1 & -1 & 1 & -1 & $R_z$ & $xy$ \\
				\( B_1 \) & 1 & 1 & -1 & -1 & $x, R_y$ & $xz$ \\
				\( B_2 \) & 1 & -1 & -1 & 1 & $y, R_x$ & $yz$ \\
			\end{tabular}
		\end{ruledtabular}
		\end{table}
	
		The Raman tensors of these modes  have the form 
		\[
		\begin{array}{cc}
		\begin{aligned}
			A_{1} &= 
			\begin{pmatrix}
				a & . & . \\
				. & b & . \\
				. & . & c
			\end{pmatrix}
		\end{aligned}
		&
		\begin{aligned}
			A_{2} &= 
			\begin{pmatrix}
				. & d & . \\
				d & . & . \\
				. & . & .
			\end{pmatrix}
		\end{aligned}
		\\
		\\[1.5em]
		\begin{aligned}
			B_{1} &= 
			\begin{pmatrix}
				. & . & e \\
				. & . & . \\
				e & . & .
				
			\end{pmatrix}
		\end{aligned}
		&
		\begin{aligned}
			B_{2} &= 
			\begin{pmatrix}
				. & . & . \\
				. & . & f \\
				. & f & .
			\end{pmatrix}
		\end{aligned}
		\end{array}
		\]
			
	        \subsection{Electronic bandstructure}\label{secelect}
\begin{figure}[t]
		\centering
		\includegraphics[width=\columnwidth]{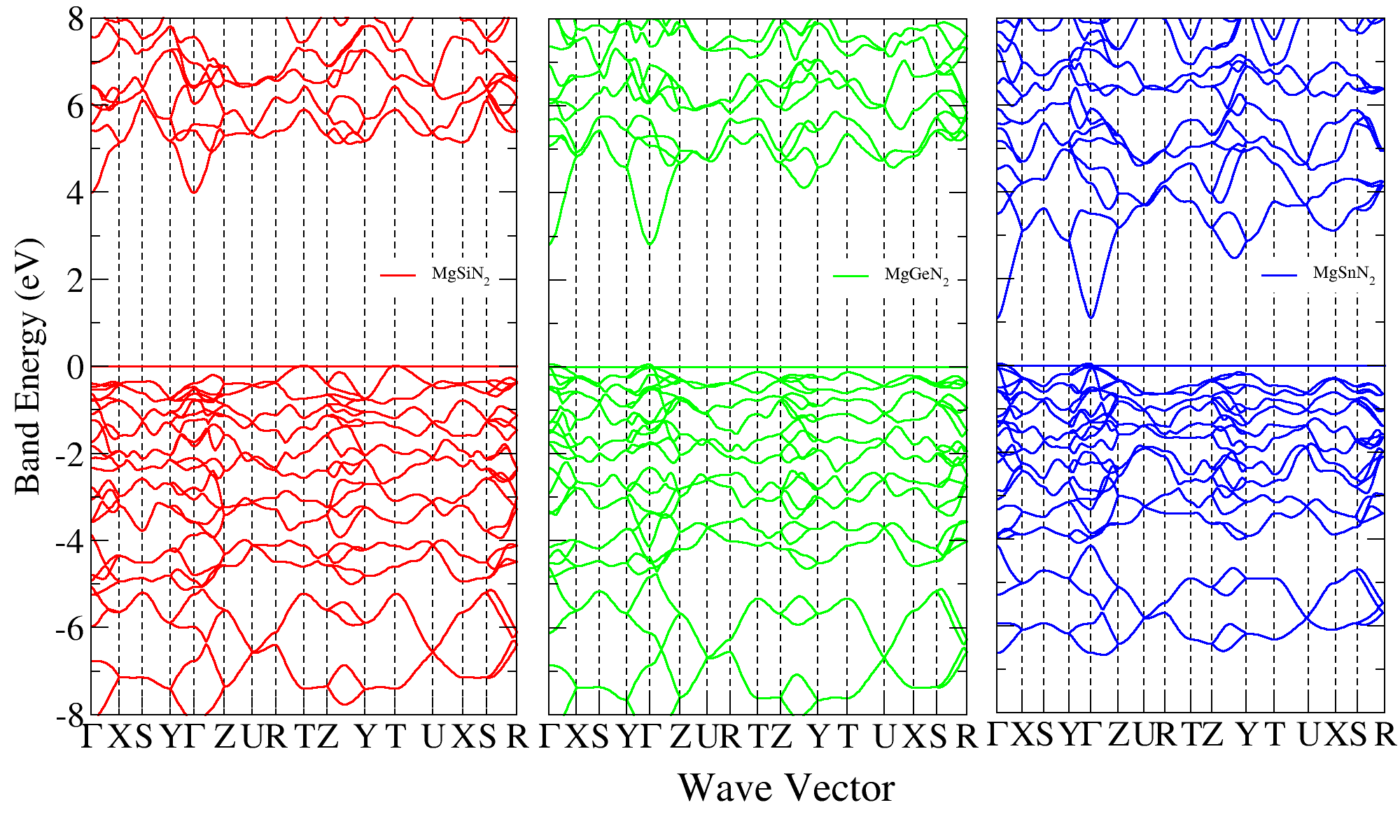}
		\caption{Electronic bandstructure of Mg-IV-N\textsubscript{2} compounds (with IV = Si,Ge,Sn) using DFT in the PBEsol exchange correlation approximation.}
		\label{bands}
		\end{figure}

                The electronic band structures in the DFT with PBE exchange correlation are shown in Fig. \ref{bands}.
                Among these three Mg-IV-N\textsubscript{2} substances, MgGeN\textsubscript{2} and MgSnN\textsubscript{2} have  a direct band gap nature at the $\Gamma$ point, whereas MgSiN\textsubscript{2} has  an indirect band gap nature with the conduction band minimum (CBM) at $\Gamma$ point and
                the valence band maximum (VBM) at the $T$ point. The Brillouin zone  symmetry points are defined as $\Gamma=(0,0,0)$, $X=(\pi/a,0,0)$, $Y=(0,\pi/b,0)$, $S=(\pi/a,\pi/b,0)$, $Z=(0,0,\pi/c)$, $U=(\pi/a,0,\pi/c)$,  $T=(0,\pi/b,\pi/c)$ and $R=(\pi/a,\pi/b,\pi/c)$. 
		
                The band gaps are compared with prior works in Table \ref{gaptable}. The band gap increases from Sn to Ge to Si. The PBE results presented here agree well with previous calculations but underestimate the gaps as usual. For a more accurate band strucrture, we refer to \cite{Atchara16,Lyu2019mg}. The band structures here are just shown to confirm the validity of our PBE
                ground state calculations which serve as starting point for the
                phonon calculation. 

                		\begin{table*}
				\caption{Electronic band gap of Mg-IV-N\textsubscript{2} compounds \label{gaptable}}
		\begin{ruledtabular}
		\begin{tabular}{cccccc}
		Material & & This Work(PBEsol)&PBE\footnote{Jarroenjittichai \etal\cite{Atchara16}} &QSGW\footnote{QS$GW$ with $0.8\Sigma$ correction from Jarroenjittichai \etal\cite{Atchara16}, and corrected by including Ge, Sn $d$ semicore electrons in paretheses from Lyu \etal \cite{Lyu2019mg}}  & Experiment \\
		\hline
		MgSiN\textsubscript{2} & Direct & 4.39 &4.44 &6.53 & 6.27-6.36 ~\cite{Hu2025-MgSi} \\
		                     & Indirect & 3.97 &4.01 & 6.08 & 5.77-5.81  \\
		MgGeN\textsubscript{2} & Direct & 2.57& 2.67 &5.36 (4.11) & 4.28 ~\cite{Hu2025-MgGe} \\
		MgSnN\textsubscript{2} & Direct & 1.05 &1.16 & 3.59 (2.28) & \\ 
		\end{tabular}
		\end{ruledtabular}
		\end{table*}

	                        \subsection{Phonon bandstructure and density of states}\label{secphband}
\begin{figure*}[htbp]
                			\centering
                			
                			\begin{minipage}{0.32\textwidth}
                				\centering
                				\includegraphics[width=\linewidth]{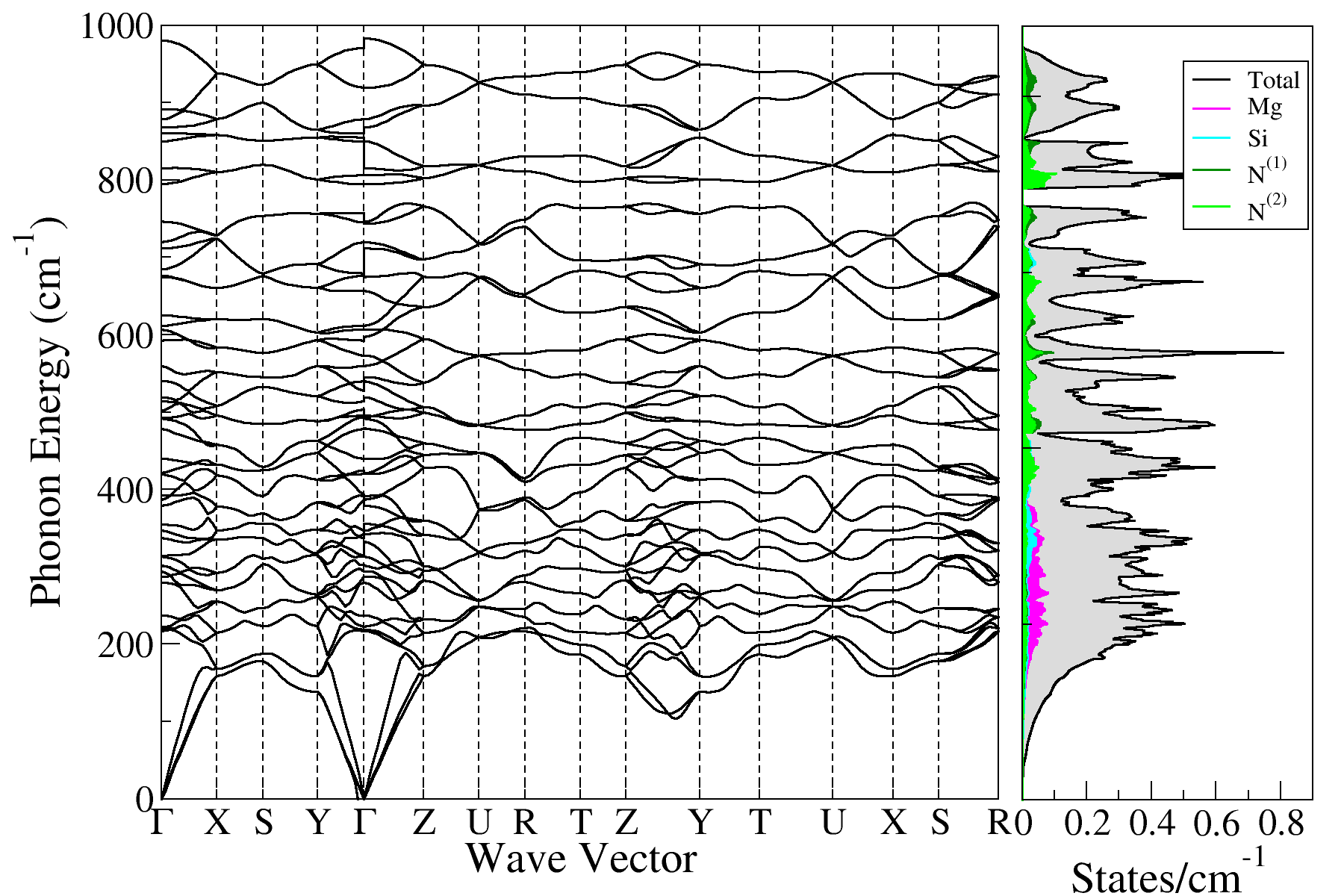}
                			\end{minipage}
                			\hfill
                			\begin{minipage}{0.32\textwidth}
                				\centering
                				\includegraphics[width=\linewidth]{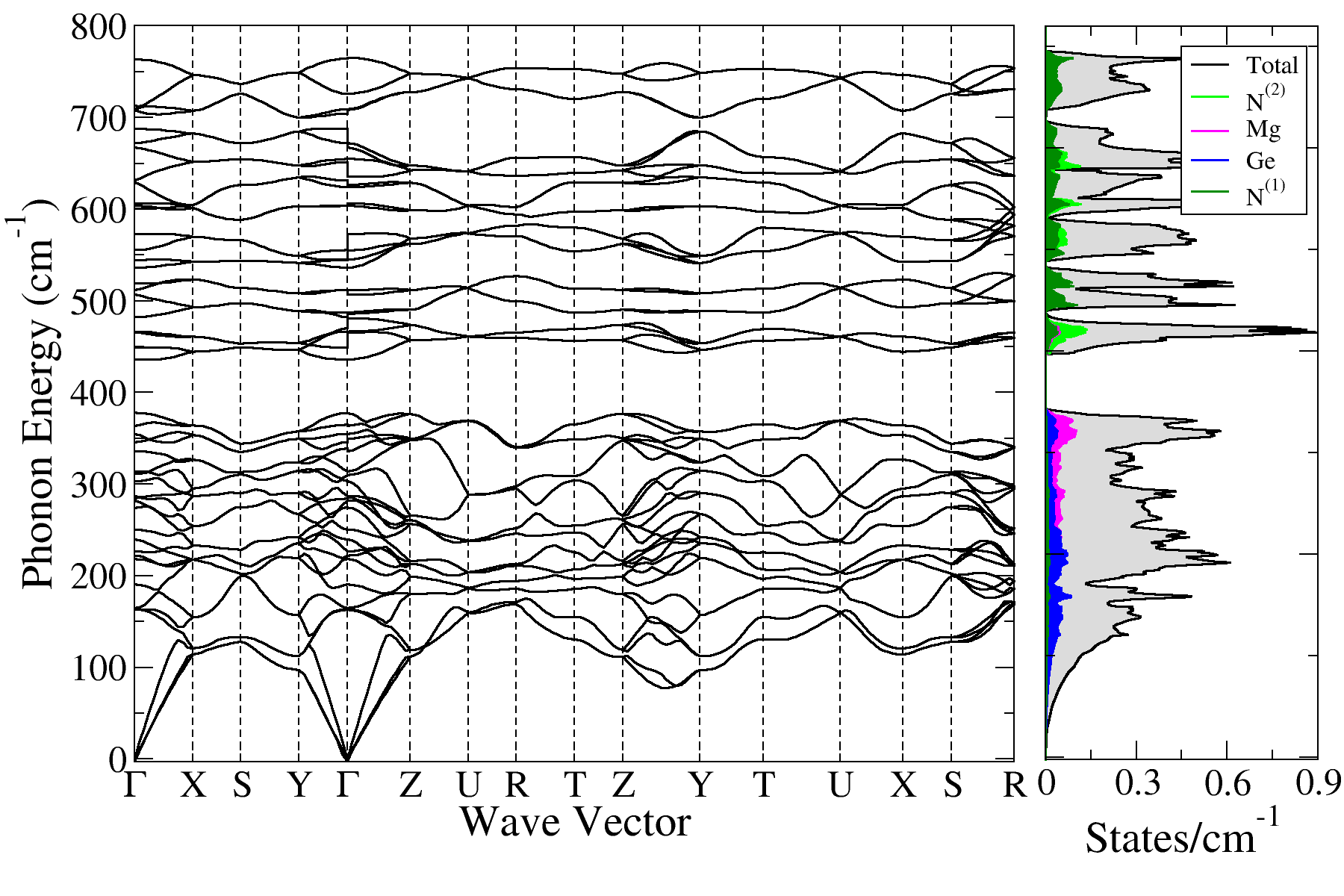}
                			\end{minipage}
                			\hfill
                			\begin{minipage}{0.32\textwidth}
                				\centering
                				\includegraphics[width=\linewidth]{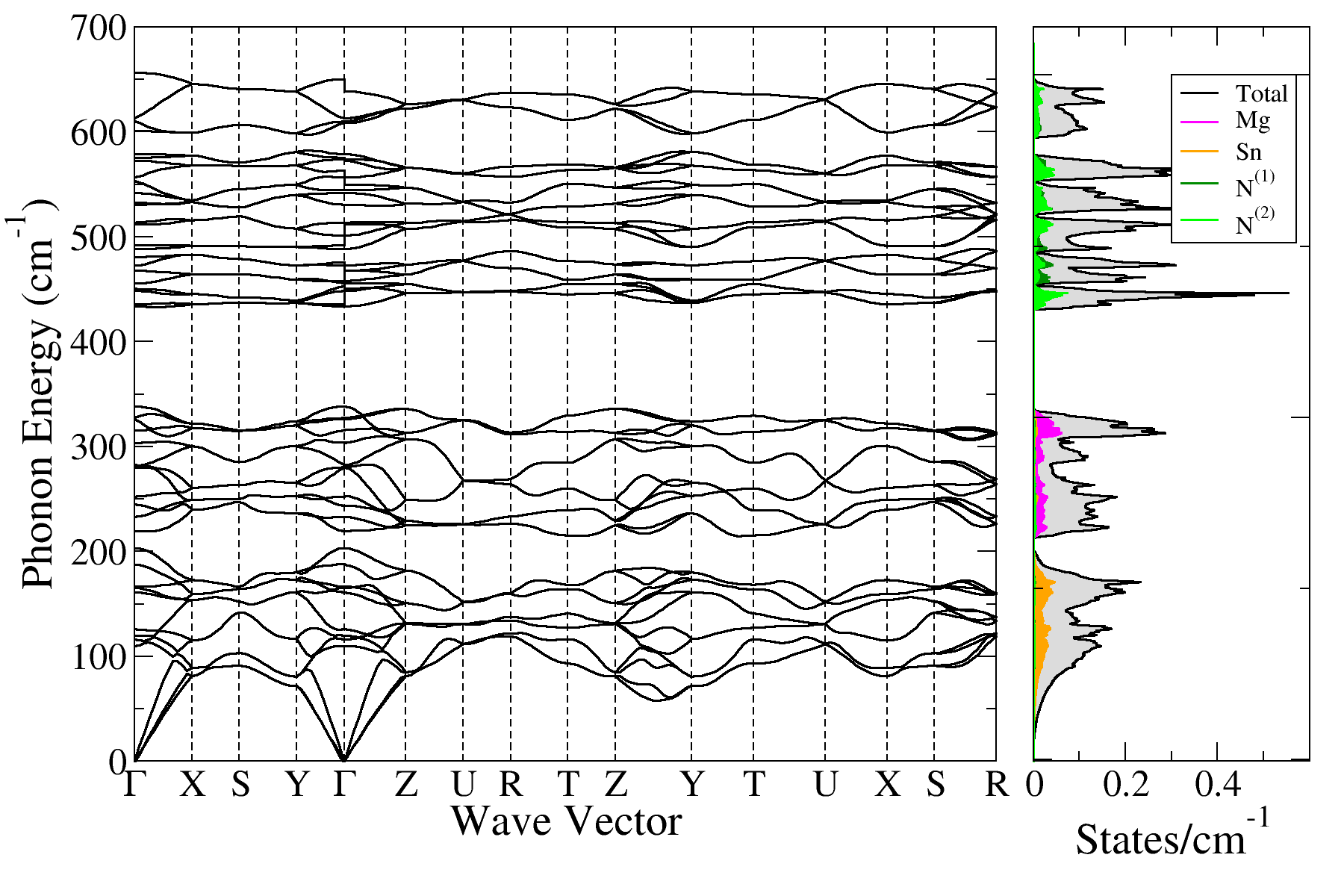}
                			\end{minipage}
                			
                			\caption{Phonon bandstructure of MgSiN\textsubscript{2} (left),
                				MgGeN\textsubscript{2} (middle), and MgSnN\textsubscript{2} (right) along the high-symmetry path and phonon densities of states indicating the contributions of one of each of the atom types.}
                			\label{phbnd}
                		\end{figure*}

                                		Phonon dispersion and atom specific phonon density of states of Mg-IV-N\textsubscript{2} compounds are shown in Fig.~\ref{phbnd}. From the phonon dispersions we can see that there are no imaginary frequencies in the system and hence we can infer that there is no mechanical instability present in these systems. We can see that the highest frequency at $\Gamma$, which sets the range of frequencies in each compound,  decreases from the  Si to the Ge to the  Sn compound. The phonon density of states decomposition in atomic contributions shows that the high frequency modes are dominated by N while the lower acoustic modes are dominated by the heavier cations, Ge and Sn, while in the MgSiN$_2$ case, the Mg  contributions are the largest at the lowest frequencies. We may note that in  MgSnN$_2$ the phonons break  into  three separate frequency ranges, the lower 12 bands are dominated by Sn (indeed 4 Sn with each 3 directions of motion gives 12 bands) and separated from the next 12 dominated by Mg and the next 24 by N. There is a large gap between the cation and N dominated bands. This is clearly a result of the large discrepancy in mass between the three types of atoms. 
The top four LO-type modes are further separated by a small gap from the lower N-optical modes. In MgGeN$_2$ we have  essentially two ranges, the cation dominated lower ones and the N dominated upper ones. 	We can still see a separation of Ge and Mg in the PDOS but the bands form one continuous range without gap for the lower 24 bands. For MgSiN$_2$, we see one continuous range of phonon bands except for the top 8 bands which are separated by a small gap from the lower ones.  While strictly speaking all but the lowest three  modes are optic, 
we can designate the lower 24 modes as "folded acoustic" type, in which  groups of atoms move together in the same direction, while in the upper 24 bands the atoms in individual bonds are moving opposite to each other. Correspondingly,  only the higher modes have a strong dipolar character and exhibit infrared absorption. 
We can further note that near $\Gamma$ the limits depend on the direction along which we approach $\Gamma$. This reflects the non-analyticity of the LO modes and the LO-TO splittings are further discussed in the next section.


	                        \subsection{Phonons at the zone center}\label{secphzone}
                              The phonon 	frequencies of the modes at $\Gamma$ labeled by their irreducible representation and their TO or LO character are given in  Table~\ref{phonon_modes}.
                              Note that because there is no inversion center, $A_1$, $B_1$ and $B_2$ modes are both infrared and Raman active while $A_2$ modes are only Raman active and have no LO-TO splitting. 
                               The $A_1$ modes have non-analytic behavior for long range electric fields along the $z$ direction and therefore LO modes with wavevector along $z$ correspond the $A_1^{LO}$ mode while 
                               along other directions correspond to the $A_1^{TO}$ mode. Likewise the $B_1^{LO}$ modes correspond to the limit of the  LO branch with wavevector along $x$ and $B_2^{LO}$ modes correspond to the LO brach with wavevector along the $y$-direction. We note however that the LO-TO splittings are small for the lower modes because they have only small dipolar character because they have in some sense still an acoustic character without strong individual bond distortions with dipolar character.

		\begin{table*}
		\caption{Phonon Modes symmetry Labeling (in cm\textsuperscript{-1} unit) \label{phonon_modes}}
		\begin{ruledtabular}
			\begin{tabular}{cccccccc}
				System & $A_{1,TO}$ & $A_{1,LO}$ & $A_{2}$ & $B_{1,TO}$ & $B_{1,LO}$ & $B_{2,TO}$ & $B_{2,LO}$ \\
				\hline
				\multirow{12}{*}{MgSiN$_2$}
				& 219.16 & 219.16 & 217.03 & 222.85 & 222.96 & 293.77 & 293.87 \\
				& 287.05 & 287.06 & 237.03 & 279.20 & 279.48 & 301.94 & 302.32 \\
				& 314.31 & 332.98 & 313.06 & 342.21 & 346.67 & 378.94 & 386.66 \\
				& 337.09 & 342.21 & 346.67 & 391.62 & 392.52 & 421.06 & 424.50 \\
				& 477.95 & 478.91 & 386.66 & 439.55 & 439.55 & 491.69 & 493.06 \\
				& 493.06 & 494.02 & 419.99 & 495.44 & 501.66 & 519.37 & 527.95 \\
				& 513.78 & 519.37 & 539.85 & 573.40 & 592.71 & 606.80 & 606.81 \\
				& 592.71 & 597.24 & 559.71 & 656.94 & 674.83 & 674.83 & 676.61 \\
				& 676.61 & 711.67 & 611.61 & 745.54 & 746.46 & 711.67 & 719.92 \\
				& 794.72 & 794.87 & 719.92 & 806.22 & 815.62 & 815.62 & 850.23 \\
				& 860.81 & 868.33 & 850.23 & 919.64 & 980.56 & 868.33 & 878.70 \\
				&   -    &   -    & 878.70 &   -    &   -    &   -    &   -    \\
				\hline 
				\multirow{12}{*}{MgGeN$_2$} 
				& 162.85 & 162.85 & 163.28 & 165.21 & 165.24 & 222.07 & 222.08 \\
				& 226.79 & 226.83 & 190.75 & 218.35 & 218.35 & 274.44 & 274.88 \\
				& 250.75 & 250.76 & 239.66 & 282.05 & 282.12 & 303.50 & 303.76 \\
				& 284.31 & 286.04 & 286.83 & 310.12 & 311.44 & 336.17 & 338.86 \\
				& 345.98 & 349.64 & 314.00 & 377.79 & 378.05 & 363.81 & 364.95 \\
				& 449.20 & 465.47 & 355.12 & 466.71 & 466.73 & 465.47 & 466.71 \\
				& 482.55 & 486.09 & 436.03 & 487.75 & 506.67 & 506.67 & 512.20 \\
				& 544.70 & 545.24 & 511.66 & 545.24 & 555.45 & 573.26 & 603.99 \\
				& 555.45 & 573.26 & 536.45 & 624.33 & 629.30 & 606.54 & 624.33 \\
				& 631.37 & 635.95 & 603.73 & 655.04 & 667.22 & 667.22 & 687.88 \\
				& 687.88 & 708.37 & 672.26 & 726.20 & 763.53 & 708.37 & 726.20 \\
				&   -    &   -    & 709.19 &   -    &   -    &   -    &   -    \\
				\hline
				\multirow{12}{*}{MgSnN$_2$}
				& 110.12 & 110.15 & 119.98 & 116.04 & 116.00 & 165.29 & 165.52 \\
				& 166.48 & 166.59 & 125.92 & 161.03 & 161.03 & 187.53 & 188.00 \\
				& 219.51 & 219.73 & 203.38 & 244.83 & 243.82 & 280.22 & 280.29 \\
				& 252.35 & 253.09 & 233.16 & 281.97 & 280.80 & 303.65 & 304.01 \\
				& 315.31 & 316.20 & 282.54 & 338.63 & 338.62 & 327.34 & 327.36 \\
				& 435.89 & 448.39 & 326.37 & 448.46 & 448.39 & 450.49 & 452.90 \\
				& 455.64 & 458.14 & 433.68 & 467.99 & 458.14 & 467.98 & 472.76 \\
				& 491.88 & 501.23 & 472.76 & 512.06 & 501.23 & 512.06 & 513.80 \\
				& 513.79 & 530.79 & 488.30 & 553.23 & 544.81 & 541.89 & 544.81 \\
				& 556.56 & 556.74 & 530.78 & 575.62 & 574.06 & 572.74 & 574.06 \\
				& 575.62 & 578.90 & 578.90 & 610.25 & 608.60 & 610.25 & 613.11 \\
				&   -    &   -    & 613.11 &   -    &   -    &   -    &   -    \\
			\end{tabular}
		\end{ruledtabular}
		\end{table*}

	        \subsection{IR Spectra and Dielectric Properties}\label{secir}
                The infrared optical properties are related to the macroscopic dielectric tensor $\varepsilon_{\alpha\beta}$ of the system defined in terms of Oscillator Strengths $S_{m,\alpha\beta}$. Which can be described by
                                \begin{equation}\label{eqn3}
                	\varepsilon_{\alpha\beta} = \varepsilon^{\infty}_{\alpha\beta} + \frac{4\pi}{\Omega_{0}} \frac{\sum_{m} S_{m,\alpha\beta}}{\omega^2_m - \omega^2 -i\Gamma_m}.
                \end{equation}
                
                The oscillator strengths in turn are obtained from the Born effective charges and eigen-mode-displacement vector, 
                \begin{equation}\label{eqn4}
                	S_{m,\alpha\beta} = \left| \sum_{\kappa,\alpha'\beta'}  Z^*_{\kappa,\alpha\beta\alpha'\beta'} U^*_{m,\textbf{q=0}}(\kappa,\alpha'\beta') \right|^2.
                \end{equation} 
                
                The Born effective charges are defined as the derivative of the macroscopic polarization $\mathscr{P}_{mac}$ with respect to the atomic displacements, or the derivative of the force on a given atom versus the long range electric field $\mathscr{E}$,  
                \begin{equation}
                	Z^\ast_{\kappa,\beta\alpha}=\Omega_0\frac{\partial\mathscr{P}_{mac,\beta}}{\partial\tau_{\kappa\alpha}(\textbf{q=0})}=\frac{\partial F_{\kappa,\alpha}}{\partial \mathscr{ E}_\beta}.
                \end{equation}
   in the {\sc Abinit } code, the changes in polarization  are  calculated from  the Berry phase in terms of the derivatives of the periodic parts of the Bloch functions as function of wave vector.           
                
                		\begin{figure}[t]
                			\centering
                			\includegraphics[width=\columnwidth]{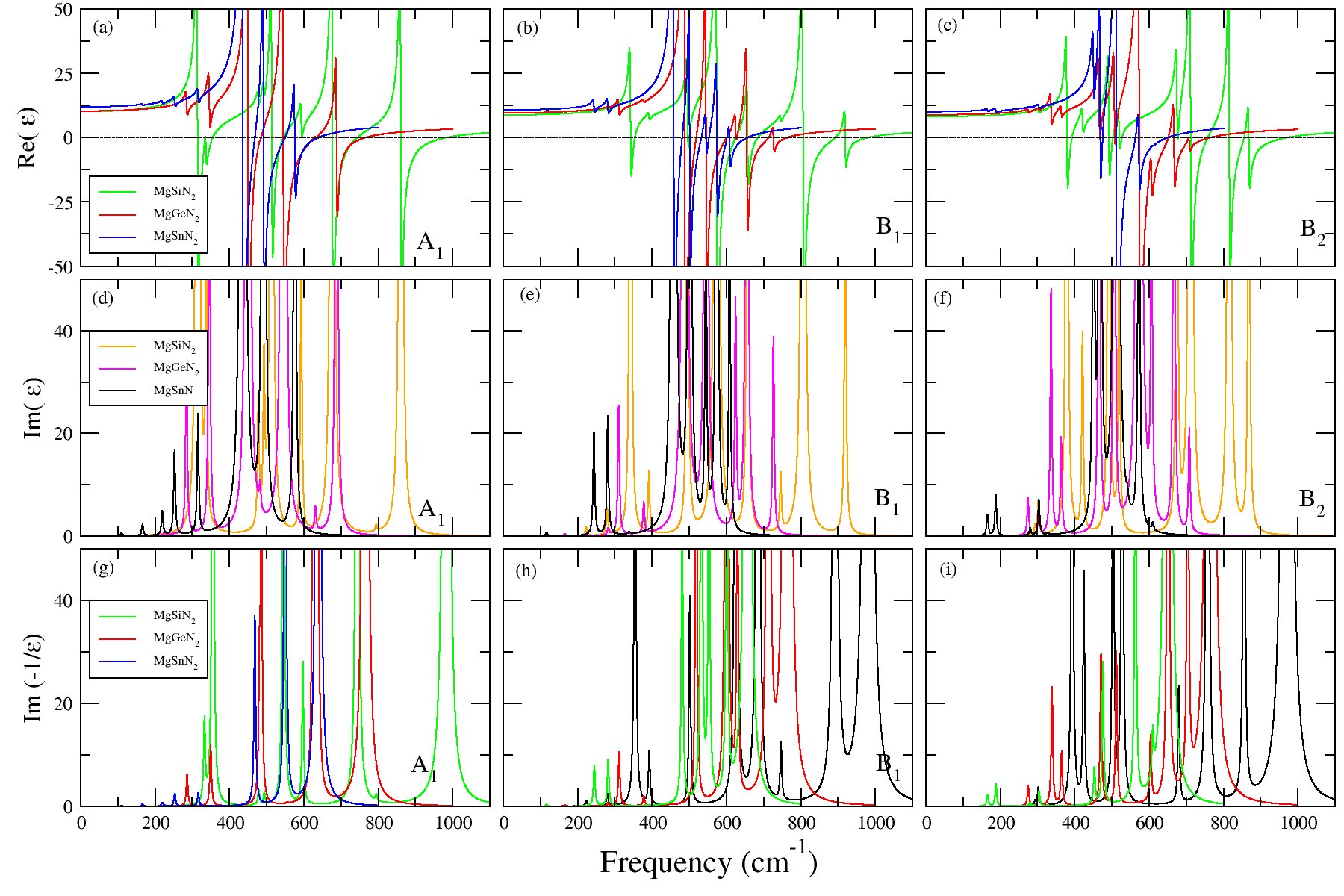}
                			\caption{Dielectric properties of Mg-IV-N\textsubscript{2} compounds (IV = Si, Ge, Sn). The first, second, and third columns correspond to A\textsubscript{1}, B\textsubscript{1}, and B\textsubscript{2} modes, respectively. The rows represent (top) real part, (middle) imaginary part, and (bottom) loss function of the dielectric response.}
                			\label{all_eps}
                		\end{figure}
            The diagonal   Born effective charge tensor components  can be found in Table.~\ref{tabborn}.  Off-diagonal terms are non-zero as well but smaller  and sum to zero among the four equivalent  atoms of any given type.  This is because the Born effective charge tensor is  a local quantity of a given atom and does not obey the symmetry restrictions of a macroscopic tensor. They indeed only have  the symmetry restrictions of the local point group of the atomic Wyckoff position, which in the present case has no symmetry at all. 
             The diagonal terms only sum to zero if we add them  over inequivalent atom types.  We can see that the Born effective charge of Mg is close to the expected value of 2  and those of the N are negative but they do not correspond the nominal valence of  $-3$ and the IV elements have values closer to $+3$ than to $+4$.  This indicates the mixed covalent/ionic type of bonding. 
                \begin{table}
			\
			\caption{Born Effective Charges. \label{tabborn}}
			\begin{ruledtabular}
				\begin{tabular}{ccccc}
					Crystal & Atom & $Z_{xx}$ & $Z_{yy}$ & $Z_{zz}$ \\ 
					\hline
					\multirow{4}{*}{MgSiN$_2$} 
					& Mg          & 1.914416  & 1.946921  & 2.079417 \\
					& Si          & 3.143441  & 3.005524  & 3.097319 \\
					& N\textsuperscript{(1)} & -2.057136 & -2.490390 & -2.996079 \\
					& N\textsuperscript{(2)} & -3.000721 & -2.462055 & -2.180658 \\
					\hline
					\multirow{4}{*}{MgGeN$_2$} 
					& Mg          & 1.859406  & 1.968732  & 2.017150 \\
					& Ge          & 3.140716  & 2.922491  & 3.116284 \\
					& N\textsuperscript{(1)} & -2.062036 & -2.463945 & -2.923244 \\
					& N\textsuperscript{(2)} & -2.938086 & -2.427278 & -2.210189 \\
					\hline
					\multirow{4}{*}{MgSnN$_2$} 
					& Mg          & 1.801440  & 1.960574  & 1.953644 \\
					& Sn          & 3.281827  & 2.989555  & 3.373626 \\
					& N\textsuperscript{(1)} & -2.193237 & -2.473043 & -2.956634 \\
					& N\textsuperscript{(2)} & -2.890030 & -2.477087 & -2.370636 \\
				\end{tabular}
			\end{ruledtabular}
		\end{table}
		
		The oscillator strengths of the Mg-IV-N\textsubscript{2} crystals are presented in the following Tables ~\ref{tabosc_si}--~\ref{tabosc_sn}.  The oscillator strengths only contain diagonal elements 
		because it is a macroscopic tensor and in a orthorhombic crystal that implies only diagonal (but still anisotropic) elements and are only non-zero for the infrared-active irreducible representations. 
		The $S_{xx}$ is only non zero for $B_1$ modes, $S_{yy}$ for $B_2$ modes and the $S_{zz}$ for $A_1$ modes. In the table the modes are numbered from low to high frequency. For example, mode 5 is the lowest $A_1$ mode because the first three modes are zero and the lowest non zero mode has $A_2$ symmetry referring back to Table \ref{phonon_modes}. We confirm that there are 11 infrared active nodes of each polarization. 
				\begin{table}
						\caption{Oscillator strenghts of MgSiN2 \label{tabosc_si}}
			\begin{ruledtabular}
				\begin{tabular}{cccc}
					
					Mode	& $S_{xx}$ & $S_{yy}$ & $S_{zz}$	\\
					\hline
					5          &  - & - & $2.5115 \times 10^{-09}$ \\
					6          &  $9.9631 \times 10^{-07}$ & - & - \\
					8          &  $3.4339 \times 10^{-06}$ & - & - \\
					9          &  - & - & $2.5697 \times 10^{-07}$ \\
					10          &  - & $1.2114 \times 10^{-06}$ & - \\
					11          &  - & $4.6916 \times 10^{-06}$ & - \\
					13          &  - & - & $4.1335 \times 10^{-04}$ \\
					14          &  - & - & $4.5114 \times 10^{-05}$ \\
					15          &  $1.6520 \times 10^{-04}$ & - & - \\
					17          &  - & $2.1704 \times 10^{-04}$ & - \\
					19          &  $1.1400 \times 10^{-05}$ & - & - \\
					21          &  - & $3.9295 \times 10^{-05}$ & - \\
					22          &  $1.1525 \times 10^{-08}$ & - & - \\
					23          &  - & - & $2.8198 \times 10^{-05}$ \\
					24          &  - & $2.2439 \times 10^{-04}$ & - \\
					25          &  - & - & $3.6226 \times 10^{-05}$ \\
					26          &  $1.6675 \times 10^{-04}$ & - & - \\
					27          &  - & - & $5.3039 \times 10^{-04}$ \\
					28          &  - & $8.3564 \times 10^{-05}$ & - \\
					31          &  $1.1403 \times 10^{-03}$ & - & - \\
					32          &  - & - & $7.6014 \times 10^{-05}$ \\
					33          &  - & $1.4502 \times 10^{-07}$ & - \\
					35          &  $2.5155 \times 10^{-04}$ & - & - \\
					36          &  - & $2.3602 \times 10^{-04}$ & - \\
					37          &  - & - & $1.0705 \times 10^{-03}$ \\
					38          &  - & $1.0833 \times 10^{-03}$ & - \\
					40          &  $2.0237 \times 10^{-05}$ & - & - \\
					41          &  - & - & $2.2933 \times 10^{-06}$ \\
					42          &  $1.2592 \times 10^{-03}$ & - & - \\
					43          &  - & $8.9755 \times 10^{-04}$ & - \\
					45          &  - & - & $1.1321 \times 10^{-03}$ \\
					46          &  - & $2.6872 \times 10^{-04}$ & - \\
					48          &  $1.9602 \times 10^{-04}$ & - & - \\
					
				\end{tabular}
			\end{ruledtabular}
			
		\end{table}
		\begin{table}
			\caption{Oscillator strenghts of MgGeN2\label{tabosc_ge}}
			\begin{ruledtabular}
				\begin{tabular}{cccc}
					Mode 	& $S_{xx}$ & $S_{yy}$ & $S_{zz}$	\\
					\hline
					4  & - & - & $1.4871 \times 10^{-08}$ \\
					6  & $1.7311 \times 10^{-07}$ & - & - \\
					8  & $3.2828 \times 10^{-08}$ & - & - \\
					9  & - & $2.1707 \times 10^{-10}$ & - \\
					10  & - & - & $5.1513 \times 10^{-07}$ \\
					12  & - & - & $1.6751 \times 10^{-07}$ \\
					13  & - & $5.4223 \times 10^{-06}$ & - \\
					14  & $9.9295 \times 10^{-07}$ & - & - \\
					15  & - & - & $2.8308 \times 10^{-05}$ \\
					17  & - & $3.6880 \times 10^{-06}$ & - \\
					18  & $2.1145 \times 10^{-05}$ & - & - \\
					20  & - & $4.3412 \times 10^{-05}$ & - \\
					21  & - & - & $8.0474 \times 10^{-05}$ \\
					23  & - & $1.8392 \times 10^{-05}$ & - \\
					24  & $6.2525 \times 10^{-06}$ & - & - \\
					26  & - & - & $1.0338 \times 10^{-03}$ \\
					27  & - & $1.6066 \times 10^{-04}$ & - \\
					28  & $1.1862 \times 10^{-06}$ & - & - \\
					29  & - & - & $6.4016 \times 10^{-06}$ \\
					30  & $1.2650 \times 10^{-03}$ & - & - \\
					31  & - & $2.0035 \times 10^{-04}$ & - \\
					34  & - & - & $1.1624 \times 10^{-03}$ \\
					35  & $7.0643 \times 10^{-04}$ & - & - \\
					36  & - & - & $2.2840 \times 10^{-05}$ \\
					37  & - & $1.7430 \times 10^{-03}$ & - \\
					39  & - & $1.1588 \times 10^{-04}$ & - \\
					40  & $7.3877 \times 10^{-05}$ & - & - \\
					41  & - & - & $7.6292 \times 10^{-06}$ \\
					42  & $5.0663 \times 10^{-04}$ & - & - \\
					43  & - & $2.3558 \times 10^{-04}$ & - \\
					45  & - & - & $4.6005 \times 10^{-04}$ \\
					46  & - & $3.8742 \times 10^{-05}$ & - \\
					48  & $7.5331 \times 10^{-05}$ & - & - \\
					
				\end{tabular}
				
			\end{ruledtabular}
		\end{table}
		
\begin{figure}[htbp]
			\centering
			\includegraphics[width=\columnwidth]{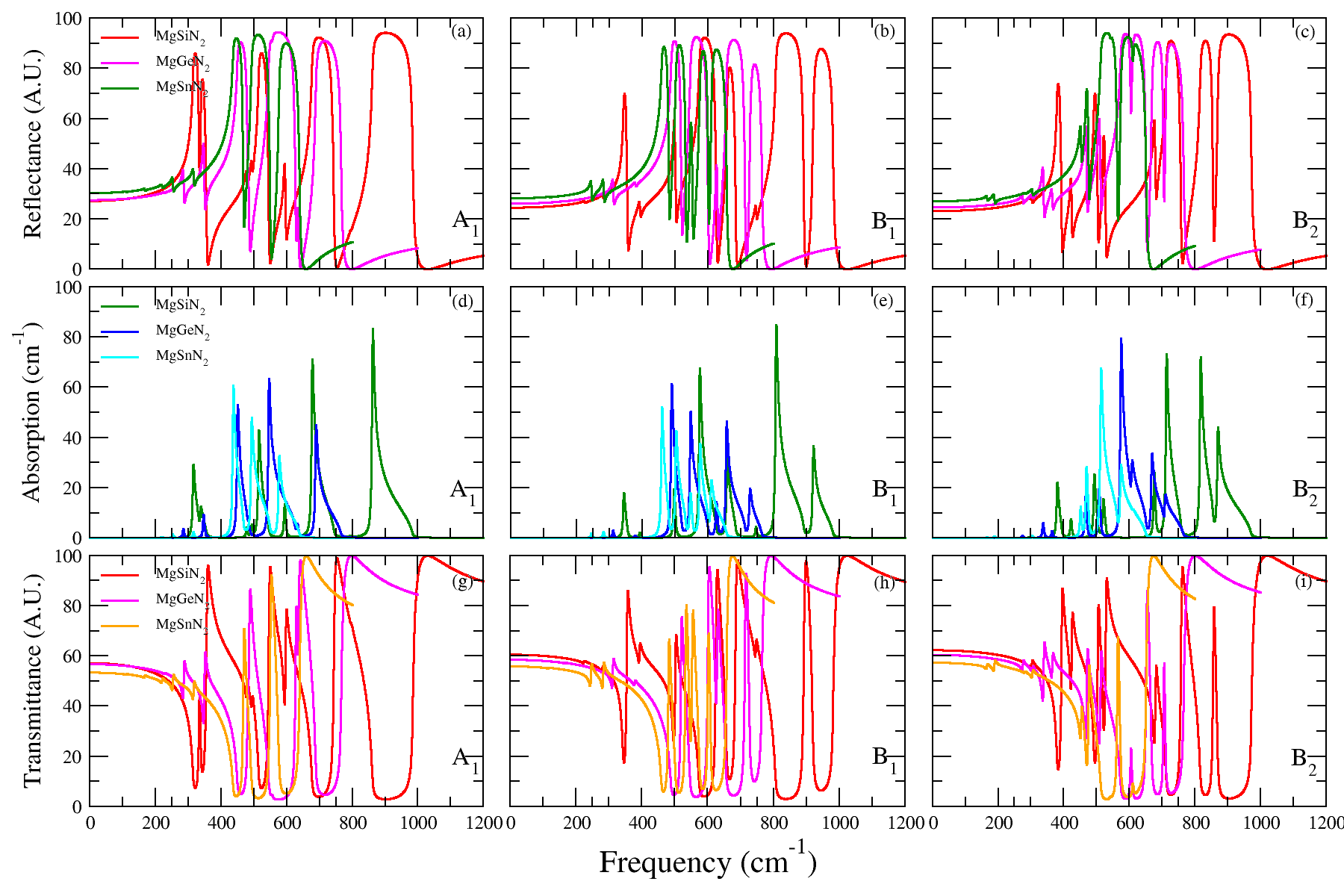}
			\caption{Macroscopic optical properties of Mg-IV-N\textsubscript{2} compounds (with IV = Si,Ge,Sn) where first columns (a),(d),(g)-represents properties corresponding to A\textsubscript{1}-modes,similarly second column(b),(e),(h) and third column (c),(f),(i) represents B\textsubscript{1} and B\textsubscript{2} modes of vibrations and their corresponding optical properties. On the other hand, first row (a),(b),(c)-represents the reflectance; second row (d),(e),(f)-represents absorption and third row (g),(h),(i)-represents the transmittance of corresponding material for corresponding mode of vibrations}
			\label{all_optical}
		\end{figure}

		\begin{table}
					\caption{Oscillator strenghts of MgSnN2 \label{tabosc_sn}}
			\begin{ruledtabular}
				\begin{tabular}{cccc}
					
					Mode 	& $S_{xx}$ & $S_{yy}$ & $S_{zz}$	\\
					\hline
					4  & - & - & $1.9939 \times 10^{-07}$ \\
					5  & $2.7431 \times 10^{-07}$ & - & - \\
					8  & $2.1582 \times 10^{-08}$ & - & - \\
					9  & - & $2.1532 \times 10^{-06}$ & - \\
					10 & - & - & $1.1532 \times 10^{-06}$ \\
					11 & - & $4.7419 \times 10^{-06}$ & - \\
					13 & - & - & $3.2749 \times 10^{-06}$ \\
					15 & $1.5619 \times 10^{-05}$ & - & - \\
					16 & - & - & $1.3434 \times 10^{-05}$ \\
					17 & - & $1.2522 \times 10^{-06}$ & - \\
					18 & $2.0722 \times 10^{-05}$ & - & - \\
					20 & - & $6.6671 \times 10^{-06}$ & - \\
					21 & - & - & $2.3414 \times 10^{-05}$ \\
					23 & - & $4.4159 \times 10^{-07}$ & - \\
					24 & $4.2512 \times 10^{-07}$ & - & - \\
					26 & - & - & $1.6293 \times 10^{-03}$ \\
					27 & $1.1141 \times 10^{-05}$ & - & - \\
					28 & - & $1.6730 \times 10^{-04}$ & - \\
					29 & - & - & $3.0983 \times 10^{-08}$ \\
					30 & $1.1895 \times 10^{-03}$ & - & - \\
					31 & - & $4.2255 \times 10^{-04}$ & - \\
					34 & - & - & $8.2788 \times 10^{-04}$ \\
					35 & $6.5495 \times 10^{-04}$ & - & - \\
					36 & - & $1.6593 \times 10^{-03}$ & - \\
					37 & - & - & $1.8540 \times 10^{-08}$ \\
					39 & - & $3.4863 \times 10^{-06}$ & - \\
					40 & $1.1643 \times 10^{-04}$ & - & - \\
					41 & - & - & $1.2417 \times 10^{-06}$ \\
					42 & - & $2.2112 \times 10^{-04}$ & - \\
					43 & $4.4147 \times 10^{-04}$ & - & - \\
					44 & - & - & $3.3168 \times 10^{-04}$ \\
					46 & $1.2491 \times 10^{-04}$ & - & - \\
					47 & - & $3.0190 \times 10^{-06}$ & - \\
				\end{tabular}
			\end{ruledtabular}
		\end{table}
		Once, the oscillator strengths and the Born effective charges are obtained we can extract the microscopic dielectric function from equation~\ref{eqn3}. The poles of the dielectric function, which correspond to peaks in its imaginary part $\varepsilon_2(\omega)$ correspond to the TO modes, while the zeros of the real part $\varepsilon_1(\omega)$ correspond to the LO modes. The latter also show up as peaks of the loss function $-\Im{[\varepsilon^{-1}(\omega)]}$. From these, one can extract the absorption coefficient $\alpha(\omega)=\omega\varepsilon_2(\omega)/n(\omega)c$ with $\tilde n(\omega)=n(\omega)+i\kappa(\omega)=\sqrt{\varepsilon(\omega)}$ the complex index of refraction and the normal incidence reflectance $R(\omega)=|\frac{\tilde{n}(\omega)-1}{\tilde{n}(\omega)+1}|$. 
		Using a fixed broadening  parameter $\Gamma_n$ of 5 cm$^{-1}$ for all modes, we obtain the dielectric function and related infrared optical properties, displayed in Fig.~\ref{all_eps} and Fig.~\ref{all_optical}. This is a typical value used to obtain a smooth spectrum. The actual phonon linewidth may vary from mode to mode and has contributions from electron-phonon and three- and four-phonon scattering anharmonic terms, and are here not calculated. When experimental spectra become available, one could adjust the linewidth to obtain a better fit but in the absence of such information, the single value used here is sufficient to provide a first approximation to the spectral shape. Some modes, having a very small oscillator strengths $\lessapprox 2\times10^{-06}$ (and small TO-LO) splitting, they are not clearly visible and hence we do not observe all of the A\textsubscript{1}, B\textsubscript{1} and B\textsubscript{2} modes in the dielectric function.
		Once we have the reflectivity and absorptions we can deduce the transmittance of each mode for each of the crystals. For a light shining on to a material of thickness $d$, it reflects $R$ amount of the photons and gets absorbed by $\alpha$ amount. Therefore, the transmittance $T = \frac{(1-R)^2 \exp^{-\alpha d}}{1 - R^2 \exp^{-2\alpha d}}$. In this case, we chose the thickness $d$ of material of amount $50$ nm as reported in the experiments~\cite{Hu2025-MgSi,Hu2025-MgGe}.

	        \subsection{Raman spectra}\label{secraman}
\begin{figure}
                	\centering
                	\includegraphics[width=\columnwidth]{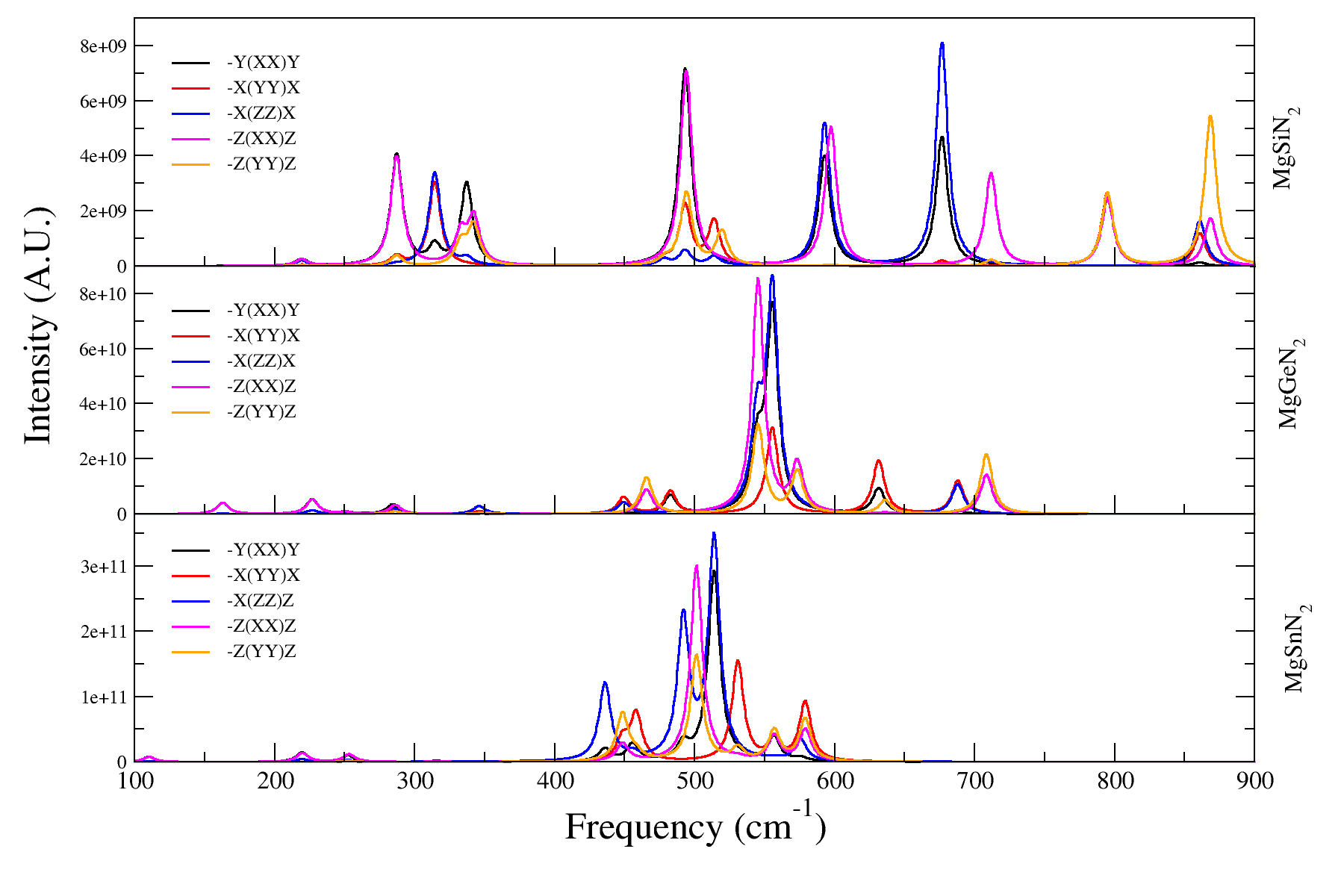}
                	\caption{Raman backscattering processes for $A_1$ modes in Mg-IV-N\textsubscript{2} compounds (with IV = Si,Ge,Sn) for different scattering geometries $k_i(e_ie_o)k_o$. The top figure shows the LO-TO splittings in MgSiN\textsubscript{2}, similarly middle figure shows MgGeN\textsubscript{2} and the bottom one represents MgSnN\textsubscript{2}}
                	\label{raman_a1}
                \end{figure}

\begin{figure}[h]
                	\centering
                	\includegraphics[width=\columnwidth]{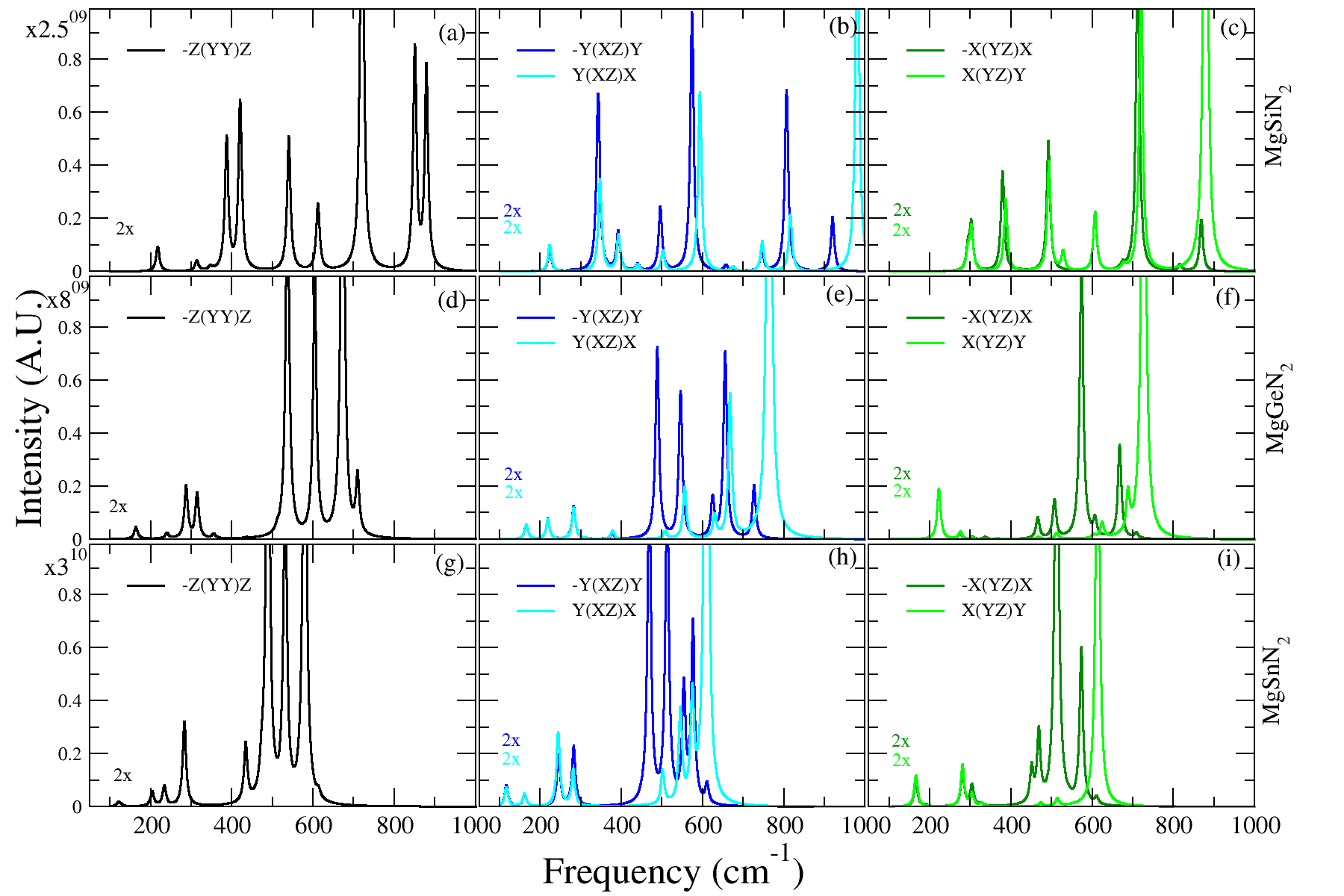}
                	\caption{Raman backscattering processes for $A_2$, $B_1$ and $B_2$ modes in Mg-IV-N\textsubscript{2} compounds (with IV = Si,Ge,Sn). LO-TO splitting is observed in $B_1$ and $B_2$ modes by comparing backscattering and right angle scattering geometries which is observed in figures (b)-(c),(e)-(f),(h)-(i).}
                	\label{raman_a2b1b2}
                \end{figure}

\begin{figure}
                	\centering
                	\includegraphics[width=\columnwidth]{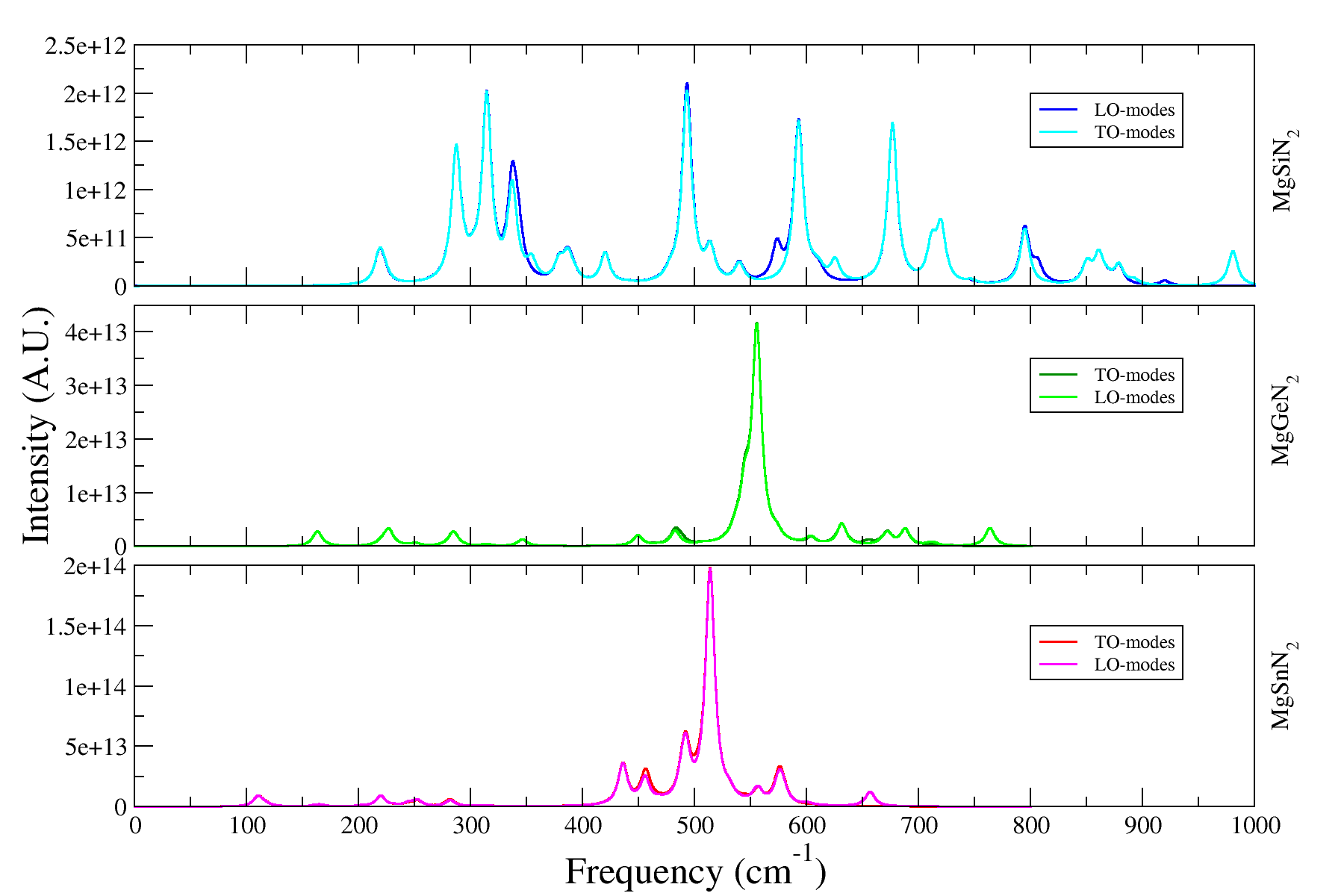}
                	\caption{This figure represents the Raman spectra of averaging over all angles for a powder sample. Top figure shows the LO-TO splittings in MgSiN\textsubscript{2}, similarly middle figure shows MgGeN\textsubscript{2} and the bottom one represents MgSnN\textsubscript{2}}
                	\label{raman_powder}
                \end{figure}

                The Raman simulated spectra are obtained by broadening the modes into a Gaussian peak with the same uniform linewidth of 5 cm$^{-1}$ as used for the IR spectra and intensity corresponding to the Raman tensor. Again, this is just a typical value and may require adjustment for a better fit to experiment. 
                In fig. \ref{raman_a1} we show the Raman spectra for $A_1$ symmetry corresponding to different backscattering geometries indicated by $k_i(e_ie_o)k_o$ with $k_i$,$k_o$ the incident and scattered wave vector direction and $e_i$,$e_o$ the incident and outgoing light polarization. The geometries with $k_i=-k_o$ along $z$ correspond to LO modes because $z$ corresponds to $A_1$ symmetry. the other cases all correspond to $A_1^{TO}$ modes.  However, because the  $A_1$ tensor has different diagonal elements for $xx$, $yy$, $zz$ polarization, the intensities are different .
                In Fig. \ref{raman_a2b1b2} we show the simulated Raman spectra for $A_2$ modes corresponding to $xy$ polarization, $B_1$ modes corresponding to $xz$ and $B_2$ modes corresponding to $yz$ polarization.  In the $B_1$ case,  backscattering geometry such as $-Y(XZ)Y$ gives the TO modes while the right angle scattering geometry $Y(XZ)X$  gives both TO and LO modes. 
                Similarly for $B_2$ $-X(YZ)X$ gives TO modes and $X(YZ)Y) $ gives both TO and LO modes. The Raman spectrum averaged over directions which is relevant for a polycrystalline sample is shown in fig.~\ref{raman_powder}.

                
                We finally note that these represent first-order Raman scattering  and assume crystal momentum conservation so that only modes at the zone center  contribute. Often in samples with significant defect induced disorder, the momentum  conservation is relaxed and one then finds a disorder induced Raman scattering which can have contributions from the whole Brillouin zone and therefore is closer to  the integrated Phonon DOS which was presented in Sec.\ref{secphband}.
                
	        \subsection{Elastic and piezoelectric properties}\label{secpiezo}
	        Piezoelectric properties of insulators can be obtained as a response of microscopic polarization, $\mathscr{P}_\alpha$, where $\alpha=\{x,y,z\}$ with respect to the homogeneous strain  $\eta_j$ with $j=\{1,...,6\}$ in Voigt notation~\cite{Nye1985} . The corresponding stress tensor is written $\sigma_i$ and the electric field components $\mathscr{E}_\alpha$ and displacement 
	        $\mathscr{D}_\alpha=\mathscr{E}_\alpha+4\pi\mathscr{P}_\alpha$. From the second diagonal second derivatives of the total energy per unit volume versus strain  and in the absence of electric field, 
	        we obtain the elastic constant tensor $C_{ij}=\partial \sigma_i/\partial \eta_j=\partial^2E/\partial \eta_i\partial\eta_j$ and its inverse the compliance tensor $S_{ij}=[C^{-1}]_{ij}=\partial\eta_i/\partial \sigma_j$. From the mixed second derivative vs. strain and electric field, we obtain the piezoelectric coefficient 
	        $e_{\alpha i}=\partial\mathscr{P}_\alpha/\partial\eta_i=\partial^2E/\partial\mathscr{E}_\alpha\partial\eta_i$.   Besides the direct piezoelectric effect, one can also define the inverse piezoelectric tensor and one may consider stress or strain  as independent variable, and $\mathscr{E}$ or $\mathscr{D}$ as independent variable.  
	        For example $e_{\alpha i}=\left(\partial{\mathscr{D}_\alpha}/\partial \eta_i\right)_\mathscr{E}=-\left(\partial\sigma_i/\partial\mathscr{E}_\alpha\right)_\eta$ and 
	        $d_{\alpha i}=\left(\partial\mathscr{D}_\alpha/\sigma_i\right)_\mathscr{E}=\left(\partial \eta_i/\partial \mathscr{E}_\alpha\right)_\sigma$., where the subscripts mean that this quantity is kept constant. 
                 It is to be understood that here the total energy per unit volume $E$ is each time minimized as function of the atomic positions in the  unit cell. If not, one would obtain the clamped ion or frozen ion versions of these tensors. A careful analysis of these definitions considering the different boundary conditions in the presence of a  finite electric field  or displacement or finite strain or stress around which the derivatives are calculated can be found in  Wu \etal~\cite{Wu_PhysRevB.72.035105} and requires replacing the total energy by an appropriate enthalpy function.

\begin{figure}[h]
		\centering
		\includegraphics[width=\columnwidth]{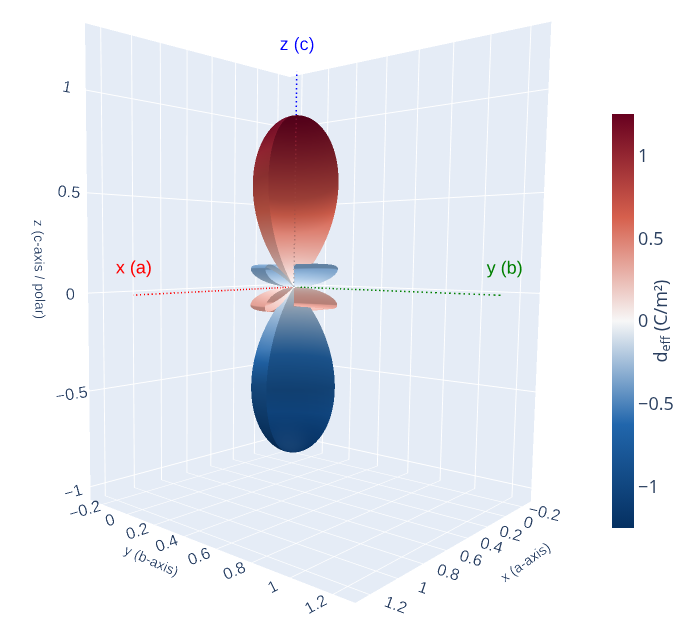}
		\caption{Longitudinal piezoelectric modulus $e(\theta,\phi)$ visualized as a spherical surface plot as function of direction for  MgSiN$_2$} 		\label{piezo1}
		\end{figure}

	        The orthorhombic symmetry $mm2$ point group of the $Pna2_1$ structure II-IV-N$_2$ compounds imposes restrictions on which components of the $e_{\alpha i}$, $d_{\alpha i}$, $C_{ij}$ and $S_{ij}$ can be non-zero.   The values for the non-zero components of the piezoelectric coefficient tensors, elastic and compliance tensors are presented in Table.~\ref{tabpiezo} and compared with the ones found in 
	        the Materials Project. The longitudinal piezoelectric modulus is defined by the polarization obtained in the same direction of a tensile strain. This can then be calculated for any direction  by rotating the tensor and plotted as function of direction. If the direction cosines of a  general direction $\hat{\bf r}$ are $l=sin{\theta}\cos{\phi},m=\sin{\theta}\sin{\phi},n=\cos{\theta}$, then in our case, the longitudinal piezoelectric modulus 
	        $e(l,m,n)=l^2n(e_{15}+e_{31})+m^2n(e_{24}+e_{32})+n^3e_{33}$ and a similar equation for $d$ and this is shown as function of direction in Fig. \ref{piezo1} for MgGeN$_2$ as a magnitude surface, in other words, in each direction we draw a vector with length the absolute value of the modulus and connect these with a surface. Note that in the Voigt notation
                $e_{\alpha i}=e_{\alpha\beta\gamma}$ if $i\in\{1,2,3\}$ and
                $e_{\alpha i}=2e_{\alpha\beta\gamma}$ if $i\in\{4,5,6\}$.
	        We plot only the region where $e(l,m,n)>0$. For the opposite z-direction we get equal magnitude but negative response.
                Note that the maximum logintudinal direction occurs for the $z$ direciton, which corresponds to the direction of the cation-nitrogen bonds which is approximately along the $c$ direciton. Clearly for this direction, we get
                the maximal piezoelecttric response because we stretch or compress the dipoles the most. Also at any $z$, the slices would be ellipses
                with major axes $e_{15}+e_{31}$ and $e_{24}+e_{32}$ which one can check are almost equal. 
			
			\begin{table*}
				\caption{Piezoelectric and elastic properties \label{tabpiezo}}
				\begin{ruledtabular}
					\begin{tabular}{cc cc cc cc}
						\multicolumn{2}{c}{Quantity} 
						& \multicolumn{2}{c}{MgSiN\textsubscript{2}} 
						& \multicolumn{2}{c}{MgGeN\textsubscript{2}} 
						& \multicolumn{2}{c}{MgSnN\textsubscript{2}} \\
						
						\multicolumn{2}{c}{} 
						& This Work & Literature~\cite{deJong2015} 
						& This Work & Literature~\cite{deJong2015} 
						& This Work & Literature~\cite{deJong2015} \\
						
						\hline
						
						\multirow{5}{*}{Piezoelectric Constant (C/m\textsuperscript{2})} 
						& e\textsubscript{15} & -0.400 & -0.427 & -0.352 & -0.385 & -0.308 & -0.408 \\
						& e\textsubscript{24} & -0.272 & -0.278 & -0.234 & -0.227 & -0.275 & -0.275 \\
						& e\textsubscript{31} & -0.524 & -0.513 & -0.419 & -0.423 & -0.433 & -0.554 \\
						& e\textsubscript{32} & -0.699 & -0.711 & -0.528 & -0.519 & -0.473 & -0.631 \\
						& e\textsubscript{33} &  1.252 & 1.312 &  1.041 & 1.073 &  1.085 & 1.280 \\
						\hline
						
						\multirow{12}{*}{Compliances ($\times10^{-2}$ TPa$^{-1}$)} 
						& s\textsubscript{11} & 4.116 & 4.000 & 4.859 & 5.000 & 5.934 & 6.000 \\
						& s\textsubscript{12} & -1.697 & -2.000 & -1.822 & -2.000 & -1.995 & -2.000 \\
						& s\textsubscript{13} & -0.293 & 0.000 & -0.599 & -1.000 & -1.188 & -1.000 \\
						& s\textsubscript{21} & -1.699 & -2.000 & -1.822 & -2.000 & -1.995 & -2.000 \\
						& s\textsubscript{22} & 4.488 & 5.000 & 5.118 & 5.000 & 6.287 & 6.000 \\
						& s\textsubscript{23} & -1.077 & -1.000 & -1.134 & -1.00 & -1.378 & -1.000 \\
						& s\textsubscript{31} & -0.292 & 0.000 & -0.599 & -1.000 & -1.188 & -1.000 \\
						& s\textsubscript{32} & -1.073 & -1.000 & -1.134 & -1.000 & -1.378 & -1.000 \\
						& s\textsubscript{33} & 3.584 & 4.000 & 4.303 & 4.000 & 5.769 & 6.000 \\
						& s\textsubscript{44} & 8.179 & 8.000 & 10.882 & 12.000 & 16.126 & 16.000 \\
						& s\textsubscript{55} & 11.512 & 12.000 & 13.657 & 16.000 & 17.312 & 20.000 \\
						& s\textsubscript{66} & 7.703 & 8.000 & 10.092 & 12.000 & 14.973 & 16.000 \\
						\hline
												
						\multirow{12}{*}{Elastic Tensor Components ($\times10^{-01}$TPa)} 
						& C\textsubscript{11} & 3.006 & 3.070 & 2.524 & 2.590 & 2.075 & 2.140 \\
						& C\textsubscript{12} & 1.290 & 1.410 & 1.037 & 1.120 & 0.794 & 0.890 \\
						& C\textsubscript{13} & 0.632 & 0.700 & 0.625 & 0.690 & 0.617 & 0.700 \\
						& C\textsubscript{21} & 1.288 & 1.410 & 1.037 & 1.120 & 0.794 & 0.890 \\
						& C\textsubscript{22} & 2.953 & 2.990 & 2.501 & 2.540 & 1.982 & 2.010 \\
						& C\textsubscript{23} & 0.989 & 1.080 & 0.804 & 0.870 & 0.637 & 0.710 \\
						& C\textsubscript{31} & 0.633 & 0.700 & 0.625 & 0.690 & 0.617 & 0.700 \\
						& C\textsubscript{32} & 0.993 & 1.080 & 0.804 & 0.870 & 0.637 & 0.710 \\
						& C\textsubscript{33} & 3.139 & 3.190 & 2.622 & 2.670 & 2.012 & 2.040 \\
						& C\textsubscript{44} & 1.222 & 1.210 & 0.918 & 0.900 & 0.620 & 0.590 \\
						& C\textsubscript{55} & 0.868 & 0.850 & 0.732 & 0.710 & 0.577 & 0.550 \\
						& C\textsubscript{66} & 1.298 & 1.290 & 0.990 & 0.980 & 0.667 & 0.640 \\
						\hline

						\multirow{5}{*}{Piezoelectric Modulus (pm/V)} 
						& d\textsubscript{15} & -4.607 &  & -4.803 &  & -5.336 &  \\
						& d\textsubscript{24} & -2.218 &  & -2.541 &  & -4.434 &  \\
						& d\textsubscript{31} & -1.335 &  & -1.701 &  & -2.917 &  \\
						& d\textsubscript{32} & -3.591 &  & -3.121 &  & -3.604 &  \\
						& d\textsubscript{33} & 5.393 &  & 5.332 &  & 7.425 &  \\
					\end{tabular}
				\end{ruledtabular}
			\end{table*}

	\section{Conclusion}
	A systematic first-principles  study has been carried out of  the lattice dynamical properties of the family  of Mg-IV-N\textsubscript{2} compounds. with group-IV elements SI, Ge and Sn. The crystal structure 
	parameters obtained here in the PBEsol exchange correlation functional are in good agreement with those reported  in the Materials Project. They indicate large distortions from the ideal wurtzite supercell, in particular with large $b/a$ and small $c/a_w$ ratios. The electronic band structures are in good agreement with prior dDFT results  but strongly underestimate the gaps compared to prior QS$GW$ results. 
	The absence of imaginary  phonon frequencies reflects the mechanical stability of these materials.  The increasing difference in atomic masses going from MgSiN$_2$ to MgGeN$_2$ to MgSnN$_2$ is shown to affect how the phonon spectrum splits in different groups of phonons separated by gaps in the phonon density of states.  In MgSnN$_2$, for example, the 12 Sn derived modes at lowest energy are separated from the 12 Mg ones and from the higher N ones, while in MgGeN$_2$, the two cation derived modes occur in one continuous range and in MgSiN$_2$ the N derived modes are continuous with the cation derived modes. 
	A  symmetry group analysis and labeling of the phonon modes has been carried out which shows that there are $11A_1$, $12A_2$, $11B_1$ and $11B_2$ modes at the $\Gamma$-point. All modes are Raman active but only the $A_1$, $B_1$, $B_2$ are infrared active and correspondingly show LO-TO splitting.  The infrared optical  spectra were calculated from the macroscopic dielectric function $\varepsilon(\omega)$ in terms of the oscillator strengths and, besides the real and imaginary parts of the latter, 
	a complete set of  derived quantities, such as  loss function, the reflectivity, the absorption coefficient and the transmission are  provided for the different polarizations of the light. as well as tables of the Born effective charges and oscillator strengths. The  Raman spectra are calculated for different scattering geometries, including backscattering and right angle scattering to provide a full account of which scattering geometries provide the modes of different symmetry  and LO or TO character. This allows one also to make a comparison between the different members of the family at a single glance. 
	The piezoelectric and elastic properties of this family of materials  are reported and found to be in good agreement with prior calculations. We also plot the piezoelectric modulus for  pure tensile strain, which gives the polarization in the $z$ direction induced by  a tensile strain, as function of direction and determine in which direction it reaches its maximum.	
	
	{\bf Data Availability} 
	The data that support the findings of this study are available within the article and at \textcolor{red}{provide a github page for the data underlying the figures}.
	
	\acknowledgements{This work was supported by the US. Air Force Research Office (AFOSR) under grant No. FA9550-22-1-0201. It made use of the High Performance Computing Resource in the Core Facility for Advanced Research Computing at Case Western Reserve University}
	
	\bibliography{aln,zgn,zsn,msn,ferroelectric,dft,piezo,piezo_data}
	
\end{document}